\documentclass[journal]{IEEEtran}

\usepackage{cite}
\usepackage{graphicx}
\usepackage{amsfonts} 
\usepackage{amsmath}
\usepackage{enumitem} 
\usepackage{bm} 

\usepackage{times}
\usepackage{psfrag,subfig,balance}
\usepackage{amssymb}
\usepackage{caption}
\usepackage{color}
\usepackage{array}
\usepackage{multirow}
\usepackage{mathtools}

\newcommand{\ts}{\textsuperscript}

\ifCLASSINFOpdf
\else
\fi
\usepackage{url}
\begin{document}
\title{The Effects of Noisy Labels on Deep Convolutional Neural Networks for Music Tagging}
\author{Keunwoo~Choi,
		Gy\"orgy~Fazekas,~\IEEEmembership{Member,~IEEE,}
        Kyunghyun~Cho,
        and~Mark~Sandler,~\IEEEmembership{Fellow,~IEEE}
\thanks{K. Choi, G. Fazekas, and M. Sandler are with
	the Centre for Digital Music, Electric Engineering and Computer Science, Queen Mary University of London, London, UK,
	e-mail: keunwoo.choi@qmul.ac.uk.}
\thanks{K. Cho is with Center for Data Science, New York University, New York, USA.}
\thanks{Manuscript received 5th June 2017; revised 10th September 2017; revised 8th October 2017}}

\markboth{Journal of \LaTeX\ Class Files,~Vol.~14, No .~8, August~2015}%
{Shell \MakeLowercase{\textit{et al.}}: Bare Demo of IEEEtran.cls for IEEE Journals}

\maketitle

\begin{abstract}
Deep neural networks (DNN) have been successfully applied to music classification including music tagging. However, there are several open questions regarding the training, evaluation, and analysis of DNNs. In this article, we investigate specific aspects of neural networks, the effects of noisy labels, to deepen our understanding of their properties. We analyse and (re-)validate a large music tagging dataset to investigate the reliability of training and evaluation. Using a trained network, we compute label vector similarities which is compared to groundtruth similarity. 

The results highlight several important aspects of music tagging and neural networks. We show that networks can be effective despite relatively large error rates in groundtruth datasets, while conjecturing that label noise can be the cause of varying tag-wise performance differences. Lastly, the analysis of our trained network provides valuable insight into the relationships between music tags. These results highlight the benefit of using data-driven methods to address automatic music tagging. 

\end{abstract}
\begin{IEEEkeywords}
Music tagging, convolutional neural networks
\end{IEEEkeywords}
\IEEEpeerreviewmaketitle


\section{Introduction}
\label{sec:intro}
\IEEEPARstart{M}{usic} tags are descriptive keywords that convey various types of high-level information about recordings such as mood (`sad', `angry', `happy'), genre (`jazz', `classical') and instrumentation (`guitar', `strings', `vocal', `instrumental') \cite{choi2016automatic}. Tags may be associated with music in the context of a folksonomy, i.e., user-defined metadata collections commonly used for instance in online streaming services, as well as personal music collection management tools. 
As opposed to expert annotation, these types of tags are deeply related to listeners' or communities' subjective perception of music. 
In the aforementioned tools and services, a range of activities including search, navigation, and recommendation may depend on the existence of tags associated with tracks. 
However, new and rarely accessed tracks often lack the tags necessary to support them, which leads to well-known problems in music information management \cite{lamere2008social}. For instance, tracks or artists residing in the long tail of popularity distributions associated with large music catalogues may have insufficient tags, therefore they are rarely recommended or accessed and tagged in online communities. This leads to a circular problem. Expert annotation is notoriously expensive and intractable for large catalogues, therefore content-based annotation is highly valuable to bootstrap these systems. Music tag prediction is often called \textit{music auto-tagging} \cite{eck2008automatic}. Content-based music tagging algorithms aim to automate this task by learning the relationship between tags and the audio content.

Music tagging can be seen as a multi-label classification problem because music can be correctly associated with more than one true label, for example, \{`rock', `guitar', `happy', and `90s'\}. This example also highlights the fact that music tagging may be seen as multiple distinct tasks. 
Because tags may be related to genres, instrumentation, mood and era, the problem may be seen as a combination of genre classification, instrument recognition, mood and era detection, and possibly others. In the following, we highlight three aspects of the task that emphasise its importance in music informatics research (MIR). 

First, collaboratively created tags reveal significant information about music consumption habits. Tag counts show how listeners label music in the real-world, which is often very different from the decision of a limited number of experts (see Section \ref{ssec:msd_tags}) \cite{saari2013semantic}. The first study on automatic music tagging proposed the use of tags to enhance music recommendation \cite{eck2008automatic} for this particular reason. 
Second, the diversity of tags and the size of tag datasets make them relevant to several MIR problems including genre classification and instrument recognition. In the context of deep learning, tags can particularly be considered a good \textit{source task} for transfer learning \cite{choi2017transfer,lee2017multi}, a method of reusing a trained neural network in a related task, after adapting the network to a smaller and more specific dataset. Since a music tagger can extract features that are relevant to different aspects of music, tasks with insufficient training data may benefit from this approach. 
Finally, investigating trained music tagging systems may contribute to our understanding of music perception and music itself. For example, analysing subjective tags such as mood and related adjectives can help building computational models for human perception of music (see Section \ref{ssec:label_vector}).

Albeit its importance, there are several issues one faces when analysing music tags. A severe problem, particularly in the context of deep learning, is the fact that sufficiently large training datasets are only available in the form of folksonomies. In these user-generated metadata collections, tags not only describe the content of the annotated items, for instance, well-defined categories such as instruments that appear on a track or the release year of a record, but also subjective qualities and personal opinions about the items \cite{Cantador2011}. Tags are often related to organisational aspects, such as self-references and personal tasks \cite{Chamberlain2016}. For instance, users of certain music streaming services frequently inject unique tags that have no significance to other users, i.e., they label music with apparently random character sequences which facilitate the creation of virtual personal collections, misappropriating this feature of the service. While tags of this nature are relatively easy to recognise and disregard using heuristics, other problems of folksonomies are not easily solved and constitute a great proportion of noise in these collections. Relevant problems include mislabelling, the use of highly subjective tags, such as those pertaining to genre or mood, as well as heterogeneity in the taxonomical organisation of tags. Researchers have been proposing to solve these problems either by imposing pre-defined classification systems on social tags \cite{Cantador2011}, or providing tag recommendation based on context to reduce tagging noise in the first place \cite{Font2014}. While the benefits of such organisation or explicit knowledge of tag categories have been shown to benefit automatic music tagging systems e.g. in \cite{Saari2016}, most available large folksonomies still consist of noisy labels.

In this paper, we do not directly address the above issues, but perform data-driven analyses instead, focussing on the effects of noisy labels on deep convolutional neural networks for automatic tagging of popular music. Label noise is unavoidable in most real-world applications, therefore it is crucial to understand its effects. We hypothesise that despite the noise, neural networks are able to learn meaningful representations that help to associate audio content with tags and show that these representations are useful even if they remain imperfect. The insights provided by our analyses may be relevant and valuable across several domains where social tags or folksonomies are used to create automatic tagging systems or in research aiming to understand social tags and tagging behaviour. The primary contributions of this paper are as follows: \textit{i)} An analysis of the largest and most commonly used public dataset for music tagging including an assessment of the distribution of labels within this dataset. \textit{ii)} We validate the groundtruth and discuss the effects of noise, e.g. mislabelling, on both training and evaluation. Finally, \textit{iii)} we provide a novel perspective on using specific network weights to analyse the trained network and obtain valuable insight into how social tags are related to music tracks. This analysis utilises parts of the weights corresponding to the final classifications. We termed these \textit{label vectors}.

The rest of the paper is organised as follows. Section~\ref{sec:background} outlines relevant problems and related works. Section~\ref{sec:dataset} presents an analysis of a large tag dataset from three different but related perspectives. First, tag co-occurrence patterns are analysed and our findings are presented in Section~\ref{ssec:msd_tags}. Second, we validate the dataset labels and discuss the effects of label noise on neural network training and evaluation in Section~\ref{ssec:valid_gt}. We then assess the capacity of the trained network to represent musical knowledge in terms of similarity between predicted labels and co-occurrences between ground truth labels in Section~\ref{ssec:label_vector}. Finally, we draw overall conclusions and discuss cross-domain applications of our methodology in Section~\ref{sec:conclusion}.
\section{Background and Related Work} \label{sec:background}

Music tagging is related to common music classification and regression problems such as genre classification and emotion prediction. The majority of prior research have focussed on extracting relevant music features and applying a conventional classifier or regressor. For example, the first auto-tagging algorithm \cite{eck2008automatic} proposed the use of mid-level audio descriptors such as Mel-Frequency Cepstral Coefficients (MFCCs) and an AdaBoost \cite{freund1996experiments} classifier. Since most audio features are extracted frame-wise, statistical aggregates such as mean, variance and percentiles are also commonly used. This is based on the assumption that the features adhere to a pre-defined or known distribution which may be characterised by these parameters. However, hand-crafted audio features do not necessarily obey known parametric distributions \cite{Baume2014,Casey2008}. Consequently, vector quantisation and clustering was proposed e.g. in \cite{hoffman2008content} as an alternative to parametric representations.

A recent trend in music tagging is the use of data-driven methods to \textit{learn} features instead of designing them, together with non-linear mappings to more compact representations relevant to the task. These approaches are often called \textit{representation learning} or \textit{deep learning}, due to the use of multiple layers in neural networks that aim to learn both low-level features and higher-level semantic categories. Convolutional Neural Networks (denoted `ConvNets' hereafter) have been providing state-of-the-art performance for music tagging in recent works \cite{dieleman2014end}, \cite{choi2016automatic}, \cite{lee2017multi}. In the rest of this section, we first review the datasets relevant to the tagging problem and highlight some issues associated with them. We then discuss the use of ConvNets in the context of music tagging.

\subsection{Music tagging datasets and their properties} \label{ssec:intro_labels}

Training a music tagger requires examples, i.e., tracks labelled by listeners, constituting a groundtruth dataset. The size of the dataset needed for creating a good tagger depends on the number of parameters in its machine learning model. Using training examples, ConvNets can learn complex, non-linear relationships between patterns observed in the input audio and high-level semantic descriptors such as generic music tags. However, these networks have a very high number of parameters and therefore require large datasets and efficient training strategies. Creating sufficiently large datasets for the general music tagging problem is difficult for several reasons. Compared to genre classification for instance, which can rely mostly on metadata gathered from services such as MusicBrainz\footnote{See \url{https://musicbrainz.org}: a crowd-sourced music meta-database.} or Last.fm\footnote{See \url{http://www.last.fm}: a personalised online radio.}, tagging often requires listening to the whole track for appropriate labelling, partly because of the diversity of tags \cite{Lorince2015}, i.e., the many different kinds of tags listeners may use or may be interested in while searching.

Tagging is often seen as an inherently ill-defined problem since it is subjective and there is an almost unconstrained number of meaningful ways to describe music. For instance, in the Million Song Dataset (MSD) \cite{bertin2011million}, one of the largest and most commonly used groundtruth sets for music tagging, there are 522,366 tags. This is outnumbering the 505,216 unique tracks present in the dataset. In fact, there is no theoretical limit on the number of labels in a tag dataset, since users often `invent' specific labels of cultural significance that cannot be found in a dictionary, yet become widely associated with niche artistic movements, styles, artists or genres. Peculiar misspellings also become commonplace and gain specific meaning, for instance, using `nu' in place of `new' in certain genre names (`nu jazz', `nu metal') suggests music with attention to pop-culture references or particular fusion styles, or the use of `grrrl' refers to bands associated with the underground feminist punk movement of the 90s. Given a degree of familiarity with the music, listeners are routinely able to associate songs with such tags, even if the metadata related to the artist or the broader cultural context surrounding a particular track is not known to them. This leads to a hypothesis underlying most auto tagging research, that is, audio can be sufficient to assign a reasonably broad range of tags to music automatically. We note that our approach, like other generic auto tagging methods, does not aim to cover the kinds of highly personal tags mentioned in Section \ref{sec:intro}.

Tags are also of different kinds and a single tag may often convey only a small part of what constitutes a good description. Tagging, therefore, is a multi-label classification problem. Consequently, the number of unique output vectors in a set increases exponentially with the number of items, while that of single-label classification only increases linearly. Given $K$ binary labels, the size of the output vector set can increase up to $2^K$. In practice, this problem is often alleviated by limiting the number of tags, usually to the top-$N$ tags given the number of music tracks a tag is associated with, or the number of users who applied them. 

The prevalence of music tags is also worth paying attention to because datasets typically exhibit an unbalanced distribution with a long-tail of rarely used tags. Regarding the diversity of the music and from the training perspective, there is an issue with non-uniform genre distributions too. In the MSD for example, the most popular tag is `rock' which is associated with 101,071 tracks. However, `jazz', the 12\ts{th} most popular tag is used for only 30,152 tracks and `classical', the 71\ts{st} popular tag is used 11,913 times only, even though these three genres are on the same hierarchical level.

\subsection{Labelling strategies} \label{ssec:labelling_strategies}

We finally have to consider a number of labelling strategies. Audio items in a dataset may be `strongly' or `weakly' labelled, which may refer to several different aspects of tagging. First, there is a question of whether only positive labels are used. A particular form of weak labelling means that only positive associations between tags and tracks are provided. This means, given a finite set of tags, a listener (or annotator) applies a tag in case s/he recognises a relation between the tag and the music. In this scenario, no negative relations are provided, and as a result, a tag being positive means it is `true', but a tag being negative, i.e. not applied, means `unknown'. The most common tags are about positiveness -- labels usually explain the existence of features, not the non-existence of them. Exceptions that describe negativeness include `instrumental', which may indicate the lack of vocals. Typical crowd-sourced datasets are weakly labelled, because it is the only practical solution to create a large dataset. Furthermore, listeners in online platforms cannot be reasonably expected to provide negative labels given the large number of possible tags. Strong labelling in this particular context would mean that disassociation between a tag and a track confirms negation, i.e., a zero element in a tag-track matrix would signify that the tag does not apply. To the best of our knowledge, \textit{CAL500} \cite{turnbull2007towards} is the biggest music tag dataset (500 songs) that is strongly labelled. Most recent research has  relied on collaboratively created, and therefore weakly-labelled datasets such as \textit{MagnaTagATune} \cite{law2009evaluation} (5,405 songs) and the \textit{MSD} \cite{bertin2011million} containing 505,216 songs if only tagged items are counted.

The second aspect of labelling relates to whether tags describe the whole track or whether they are only associated with a segment where a tag is considered to be true. Time-varying annotation is particularly difficult and error prone for human listeners, therefore it does not scale. Multiple tags may be applied on a fixed-length segment basis, as is done in smaller datasets such as MagnaTagATune for 30s segments. The MSD uses only track-level annotation, which can be considered a form of weak labelling. From the perspective of training, this strategy is less adverse in case of particular tags than it is for some others. Genre or era tags are certainly more likely to apply to the whole track consistently than instrument tags for instance. This discrepancy may constitute noise in the training data. Additionally, often only preview clips are available to researchers. This forces them to assume that tags are correct within the preview clip too, which constitutes another source of groundtruth noise. In this work, we train ConvNets to learn the association between track-level labels and audio recordings using preview clips associated with the MSD.

Learning from noisy labels is an important problem, therefore several studies address this in the context of conventional classifiers such as support vector machines \cite{frenay2014classification}. In deep learning research, \cite{mnih2012learning} assumes a binary classification problem while \cite{sukhbaatar2014learning} deals with multi-class classification. Usually, auxiliary layers are added to learn to fix the incorrect labels, which often requires a noise model and/or an additional clean dataset. Both solutions, together with much other research, are designed for single-class classifications, and there is no existing method that can be applied for music tagging when considered as multi-label classification and when the noise is highly skewed to negative labels. This will be discussed in Section \ref{ssec:valid_gt}.

\subsection{Convolutional neural networks}\label{ssec:cnn}
ConvNets are a special type of neural network introduced in computer vision to simulate the behaviour of the human vision system \cite{lecun1998gradient}. ConvNets have convolutional layers, each of which consists of convolutional kernels. The convolutional kernels sweep over the inputs, resulting in weight sharing that greatly reduces the number of parameters compared to conventional layers that do not sweep and are fully-connected instead. Kernels are trained to find and capture local patterns that are relevant to the task using error backpropagation and gradient descent algorithms.  
Researchers in music informatics are increasingly taking advantage of deep learning techniques. ConvNets have already been used for chord recognition \cite{humphrey2012rethinking}, genre classification \cite{li2010audio}, onset detection \cite{schluter2013musical}, music recommendation \cite{van2013deep}, instrument recognition \cite{han2017deep} and music tagging \cite{choi2016automatic}, \cite{dieleman2014end}, \cite{lee2017multi}. 

In MIR, the majority of works use two dimensional time-frequency representations as inputs, e.g., short-time Fourier transform or mel-spectrograms \cite{choi2017tutorial}.
Recently, several works proposed learning 2D representations by applying one dimensional convolution to the raw audio signal \cite{dieleman2014end,lee2017sample}. It is possible to improve performances by learning more effective representations, although the approach requires increasingly more data, which is not always available. Moreover, these approaches have been shown to learn representations that are similar to conventional time-frequency representations that are cheaper to compute \cite{dieleman2014end}, \cite{lee2017sample}. 

ConvNets have been applied to various music and audio related tasks, assuming that certain relevant patterns can be detected or recognised by cascaded one- or two dimensional convolutions. They provide state-of-the-art performance in several music information retrieval tasks including music segmentation \cite{ullrich2014boundary}, beat detection \cite{bock2016joint} and tagging \cite{lee2017multi}, as well as in non-music tasks such as acoustic event detection \cite{jansen2017large}. 
There are several possible arguments to justify the use of ConvNets for music tagging. First, music tags are often considered among the topmost high-level features representing song-level information above mid-level or intermediate musical features such as chords, beats, and tonality. This hierarchy fits well with ConvNets as they can learn hierarchical features over multilayer structures. Second, the invariance properties of ConvNets such as translation, distortion and local invariances can be useful for learning musical features when the relevant feature can appear at any time or frequency range with small time and frequency variances. 

There are many different architectures for music tagging, but many share a common training scheme. They follow the supervised learning framework with backpropagation and stochastic gradient descent, they regard the problem as a regression problem. Many of them also use cross-entropy or mean square error as a loss function, which is empirically minimised using a training set with the maximum likelihood approach. The analyses presented in this paper aim at understanding the behaviour of ConvNets using supervised learning with noisy labels. This aspect of the research is tangential to the variations of ConvNet structures. Therefore, we omit results related to the different possible ConvNet structures.
Particularly with respect to the analysis of the effect of label noise on tagging performance, a major contribution of this paper, different convnet structures have previously shown an almost identical trend in tag-wise performances \cite{choi2017convolutional}.

\subsection{Evaluation of tagging algorithms}
There are several methods to evaluate tagging algorithms. Since the target is typically binarised to represent if the $i$\ts{th} tag is true or false ($y_i \in \{0, 1\}$), classification evaluation metrics such as `Precision' and `Recall' can be used if the prediction is also binarised. Because label noise is mostly associated with negative labels, as we quantify in Section \ref{ssec:valid_gt}, using recall is appropriate since it ignores incorrect negative labels. We have to note that metrics such as recall cannot be used as a loss function since they are not differentiable. They can be used instead as an auxiliary method of assessment after training. This strategy can work well because it prevents the network from learning trivial solutions for those metrics. For instance, predicting all labels to be `True' to obtain a perfect recall score. Optimal thresholding for binarised prediction is an additional challenge however and discards information. The network learns a maximum likelihood solution with respect to the training data, which is heavily imbalanced, therefore the threshold should be chosen specifically for each tag. This introduces an entirely new research problem which we do not address here.
The area under curve - receiver operating characteristic (AUC-ROC, or simply AUC) works without binarisation of predictions and is often used as an evaluation metric. A ROC curve is created by plotting the true positive rate against the false positive rate. As both rates range between $[0, 1]$, the area under the curve also ranges between $[0, 1]$. However, the effective range of AUC is $[0.5, 1]$ since random classification yields $0.5$ when the true positive rate increases at the exact same rate of false positives.

\section{Experiments and Discussions}\label{sec:dataset}

In this section, we present the methods and the results of experiments that analyse the Million Song Dataset (MSD) and a network trained on it. We select the MSD as it is the largest public dataset available for training music taggers. It also provides crawlable track identifiers for audio signals, which enables us to access the audio and re-validate the tags manually by listening. The analyses are divided into three parts and discussed separately in the following subsections. Section \ref{ssec:msd_tags} is concerned with mutual relationships between tags. In Section \ref{ssec:valid_gt}, we re-validate the groundtruth of the dataset to ascertain the reliability of research that uses it. Section \ref{ssec:label_vector} discuss properties of the trained network.

\begin{table}[t]
	\centering
	\caption{Details of the \texttt{compact-convnet} architecture. 2-dimensional convolutional layer is specified by (channel, (kernel lengths in frequency, time)). Pooling layer is specified by (pooling length in frequency and time). Batch normalization layer \cite{ioffe2015batch} and exponential linear unit activation \cite{clevert2015fast} are used after all convolutional layers.}
	\begin{tabular}{|l|c|}
		\hline
		\multicolumn{2}{|c|}{\textit{\textbf{input (1, 96, 1360)}}} \\ \hline
		Conv2d and Max-Pooling         & (32, (3, 3)) and (2, 4)                 \\ \hline
        \multicolumn{2}{|c|}{Batch normalization - ELU activation} \\ \hline
		Conv2d and Max-Pooling         & (32, (3, 3)) and (4, 4)                 \\ \hline
        \multicolumn{2}{|c|}{Batch normalization - ELU activation} \\ \hline
		Conv2d and Max-Pooling         & (32, (3, 3)) and (4, 5)                 \\ \hline
        \multicolumn{2}{|c|}{Batch normalization - ELU activation} \\ \hline
		Conv2d and Max-Pooling         & (32, (3, 3)) and (2, 4)                 \\ \hline
        \multicolumn{2}{|c|}{Batch normalization - ELU activation} \\ \hline
		Conv2d and Max-Pooling         & (32, (3, 3)) and (4, 4)                 \\ \hline
        \multicolumn{2}{|c|}{Batch normalization - ELU activation} \\ \hline
		Fully-connected layer         & (50)                                    \\ \hline
		\multicolumn{2}{|c|}{\textit{\textbf{output (50)}}}  \\ \hline
	\end{tabular}
	\label{table:c_ConvNet}
\end{table} 

The tags in the MSD are collected using the Last.fm API which provides access to crowd-sourced music tags. We use the top 50 tags sorted by popularity (occurrence counts) in the dataset. The tags include genres (`rock', `pop', `jazz', `funk'), eras (`60s' -- `00s') and moods (`sad', `happy', `chill'). There are 242,842 clips with at least one of the top 50 tags.
The tag counts range from 52,944 (`rock') to 1,257 (`happy') and there are 12,348 unique tag vectors represented as a joint variable of 50 binary values.

Throughout this paper, particularly in Section \ref{sssec:fx_training}, \ref{sssec:val_eval}, and \ref{ssec:label_vector}, we use a ConvNet named `\texttt{compact-convnet}'. As mentioned earlier, the proposed analysis is structure-agnostic, and we chose this network since it achieves a reasonable performance while being easy to understand and analyse due to its simple structure. Table \ref{table:c_ConvNet} summarises the hyperparameters which are similar to the network in \cite{choi2016automatic}. The original audio files are encoded in mp3 format with a sampling rate of 22,050 Hz and 64 kbps constant bit-rate. They are decoded, down-mixed to monaural, re-sampled to 12 kHz, and converted into mel-spectrograms with 96 mel-bins through a windowed short-time Fourier transform using 512-point FFT with 50\% overlap. The ConvNet consists of homogeneous 5-layer, $3\times 3$ convolutional kernels. On the input side, the mel-spectrogram magnitude is mapped using decibel scaling ($\log_{10} \textbf{X}$) and adjusted using track-wise zero-mean unit-variance standardisation. 
We use 201,672\hspace{0.03cm}/\hspace{0.03cm}12,633\hspace{0.03cm}/\hspace{0.03cm}28,537 tracks as training\hspace{0.03cm}/\hspace{0.03cm}validation\hspace{0.03cm}/\hspace{0.03cm}test sets respectively, following the set splits provided by the MSD.
This network\footnote{The trained network and split settings are provided online: \\\url{https://github.com/keunwoochoi/transfer_learning_music} and \\\url{https://github.com/keunwoochoi/MSD_split_for_tagging}} achieves an AUC of 0.845.

\subsection{Tag co-occurrences in the MSD}\label{ssec:msd_tags}
We investigate the distribution and mutual relationships of tags in the dataset. This procedure helps understanding the task. Furthermore, our analysis represents information embedded in the training data. This will be compared to knowledge we can extract from the trained network (see Section \ref{ssec:label_vector}). 

\begin{figure}[t]
	\centering
	\includegraphics[width=1.0\columnwidth]{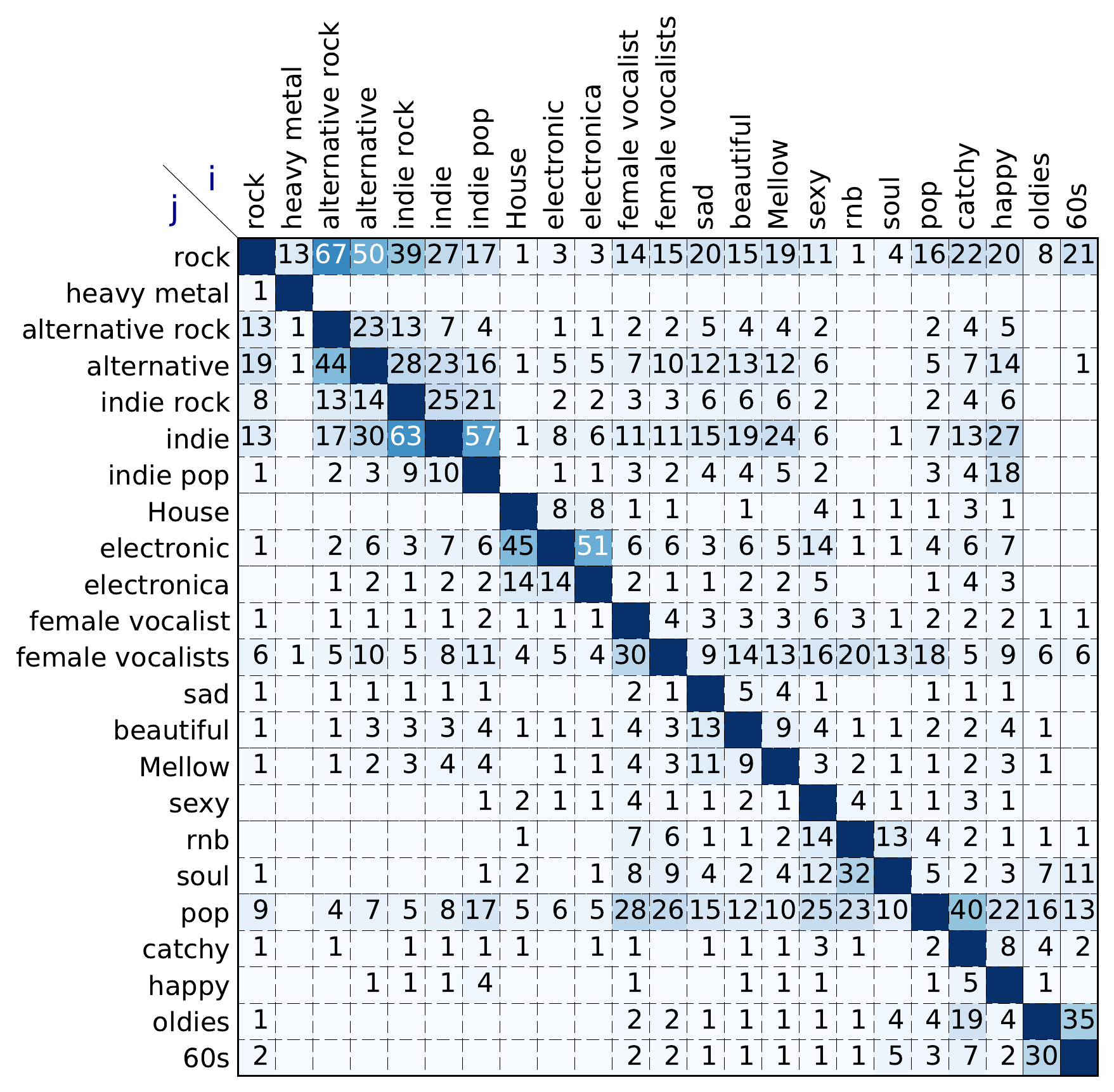}
	\caption{Normalised tag co-occurrence pattern of the selected 23 tags from the training data. For the sake of visualisation, we selected 23 tags out of 50 that have high co-occurrences and represent different categories; genres, instruments and moods. The values are computed using Eq. \ref{eq:com} (and are multiplied by 100, i.e., shown in percentage), where $y_i$ and $y_j$ respectively indicate the labels on the x-axis and y-axis.}
	\label{fig:tag_cooccur} 
\end{figure}

Here, we investigate the tuple-wise\footnote{These are not pairwise relations since there is no commutativity due to the normalisation term.} relations between tags and plot the resulting normalised co-occurrence matrix (NCO) denoted $\mathbf{C}$. Let us define \#$y_i \coloneqq |\{(x,y) \in D | y = y_i\}|$, the total number of the data points with $i$\ts{th} label being \textit{True} given the dataset $D$ where $(x, y)$ is an (input, target) pair. In the same manner, \#$(y_i \land y_j)$ is defined as the number of data points with both $i$\ts{th} and $j$\ts{th} labels being \textit{True}, i.e., those two tags \textit{co-occur}. NCO is computed using Eq. \ref{eq:com} and illustrated in Fig. \ref{fig:tag_cooccur}.
\begin{equation} \label{eq:com}
   C(i, j)= \# (y_i\land y_j)/ \#y_i.
\end{equation}

In Fig.\ref{fig:tag_cooccur}, the $x$- and $y$-axes correspond to $i, j$ respectively. Note that $C(i, j)$ is not symmetric,
e.g., (`alternative rock', `rock')~$=\#\text{(alternative rock} \land \text{rock)}/\#\text{alternative rock}$.

These patterns reveal mutual tag relationships which we categorise into three types: \textit{i)} tags belonging to a genre hierarchy, \textit{ii)} synonyms, i.e., semantically similar words and \textit{iii)} musical similarity. Genre hierarchy tuples include for instance (`alternative rock', `rock'), (`house', `electronic'), and (`heavy metal', `metal'). All first labels are sub-genres of the second. Naturally, we can observe that similar tags such as (`electronica', `electronic') are highly likely to co-occur. Lastly, we notice tuples with similar meaning from a musical perspective including (`catchy', `pop'), (`60s', `oldies'), and (`rnb', `soul').
Interestingly, $C(i, j)$ values with highly similar tag pairs, including certain subgenre-genre pairs, $y_i$ and $y_j$ are not close to 100\% as one might expect. For example the pairs (`female vocalist', `female vocalists') and (`alternative', `alternative rock') reach only 30\% and 44\% co-occurrence values, while the pairs (`rock' and `alternative rock'), (`rock', `indie rock') reach only 69\% and 39\% respectively. This is primarily because \textit{i)} items are weakly labelled and \textit{ii)} there is often a more preferred tag to describe a certain aspect of a track compared to others. For instance, `female vocalists' appears to be preferred over `female vocalist' in our data as also noted in \cite{lamere2008social}.
The analysis also reveals that certain types of label noise related to missing tags or taxonomical heterogeneity turn out to be very high in some cases. For instance, only 39\% of `indie rock' tracks are also tagged `rock'.
The effect of such label noise is studied more deeply in Section \ref{ssec:valid_gt}. Furthermore, the computed NCO under-represents these co-occurring patterns. This effect is discussed in Section \ref{ssec:label_vector}.

\subsection{Validation of the MSD as groundtruth for auto-tagging} \label{ssec:valid_gt}

\begin{table*}[t]
	\centering
	\caption{The scores of groundtruth with respect to our strongly-labelled manual annotation (subset100) in (a)-(d) and occurrence counts by the groundtruth (e), estimation (f), and on Subset400.}
	\label{table:msd_stats}
	\begin{tabular}{l|r|r|r|r||r|r|r}
		                & \multicolumn{4}{c||}{\begin{tabular}[c]{@{}l@{}} Scores of the groundtruth on Subset100\\  \end{tabular}}                                                                   & \multicolumn{3}{c}{Occurrence counts}     \\ \hline
		& \multicolumn{1}{l|}{\begin{tabular}[c]{@{}c@{}}(a)\\Error rate,\\Positive label $[\text{\%}]$\end{tabular}} & \multicolumn{1}{l|}{\begin{tabular}[c]{@{}c@{}}(b)\\Error rate,\\Negative label $[\text{\%}]$\end{tabular}} & \multicolumn{1}{c|}{\begin{tabular}[c]{@{}c@{}}(c)\\Precision\\ $[\text{\%}]$\end{tabular}} & \multicolumn{1}{c||}{\begin{tabular}[c]{@{}c@{}}(d)\\Recall\\$[\text{\%}]$\end{tabular}} & \multicolumn{1}{c|}{\begin{tabular}[c]{@{}c@{}}(e)\\In groundtruth\\(for all items)\end{tabular}} & \multicolumn{1}{c|}{\begin{tabular}[c]{@{}c@{}}(f)\\Estimate by Eq.\ref{eq:est_count}\\and Subset100\end{tabular}} & \multicolumn{1}{c}{\begin{tabular}[c]{@{}c@{}}(g)\\By our annotation\\(on Subset400)\end{tabular}} \\ \hline
		instrumental     & 6.0  & 12.0  & 94.0  & 88.7  & 8,424  (3.5\%)& 36,048 (14.9\%)&  85 (21.3\%) \\ \hline
        female vocalists & 4.0  & 24.0  & 96.0  & 80.0  & 17,840 (7.3\%) & 71,127 (29.3\%) &  94 (23.5\%)  \\ \hline
		male vocalists   & 2.0  & 64.0  & 98.0  & 60.5  & 3,026  (1.2\%)& 156,448 (64.4\%)&   252 (64.0\%) \\ \hline
		guitar           & 2.0  & 70.0  & 98.0  & 58.3  & 3,311  (1.4\%)& 170,916 (70.4\%)&  266 (66.5\%) \\  
	\end{tabular}
\end{table*}

Next, we analyse the groundtruth noise in the MSD and examine its effect on training and evaluation. There are many sources of noise including incorrect annotation as well as information loss due to the trimming of full tracks to preview clips. Some of these factors may be assumed to be less adverse than others. In large-scale tag datasets, the frequently used weak labelling strategy (see Section\ref{ssec:labelling_strategies}) may introduce a significant amount of noise. This is because by the definition of weak labelling, considering numerous tags a large portion of items remain unlabelled, but then these relations are assumed to be negative during training. 

Validation of the annotation requires re-annotating the tags after listening to the excerpts, which is not a trivial task for several reasons. First, manual annotation does not scale and requires significant time and effort. Second, there is no single correct answer for many labels -- music genre is an ambiguous and idiosyncratic concept, emotion annotation is highly subjective too, so as labels such as `beautiful' or `catchy'. Instrumentation labels can be objective to some extent, assuming the annotators have expertise in music. Therefore, we re-annotate items in two subsets using four instrument labels as described below. 

\begin{itemize}
\item \textbf{Labels:} `instrumental', `female vocalists', `male vocalists',  `guitar'.
\item \textbf{Subsets:}
  \begin{itemize}
    \renewcommand{\labelitemii}{\tiny$\blacksquare$}
    \item {Subset100:} randomly selected 100 items for each class. All are from the training set and positive/negative labels are balanced as 50\hspace{0.03cm}/\hspace{0.03cm}50 respectively.
    \item {Subset400:} randomly selected 400 items from the test set.
  \end{itemize}
\end{itemize}

\begin{figure}[t]
	\centering
	\includegraphics[width=1.0\linewidth]{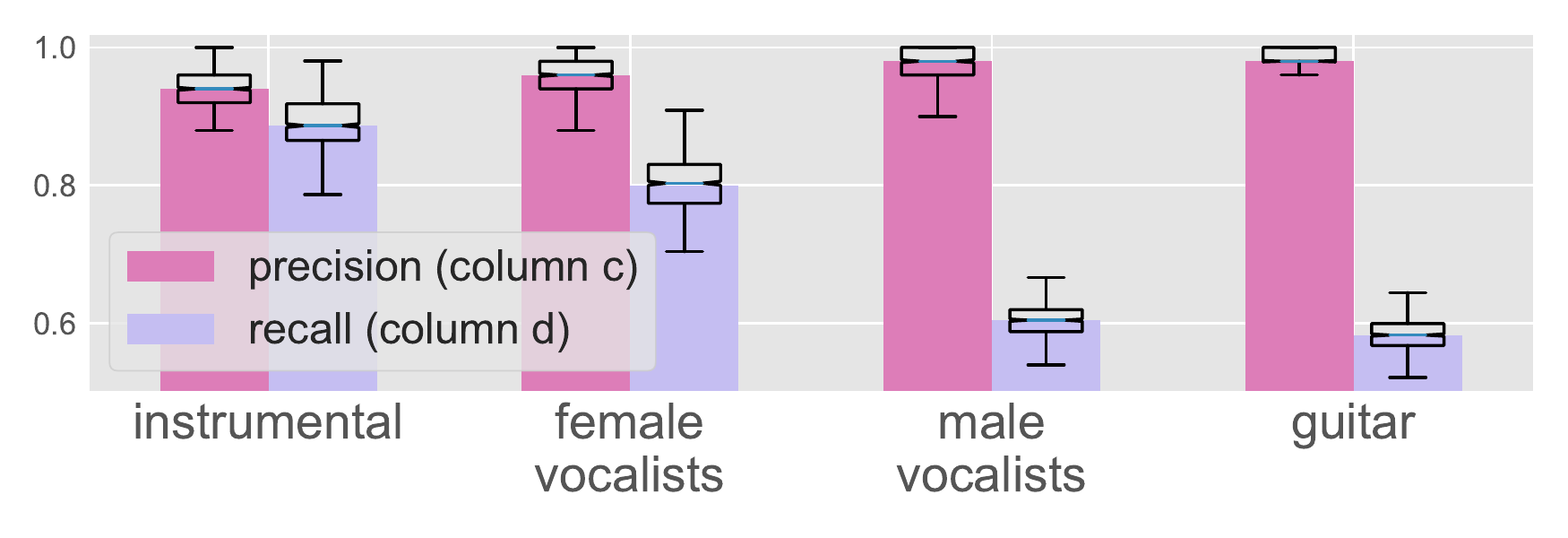}
	\caption{The precision and recall of groundtruth on Subset100, corresponding to columns (c) and (d) in Table \ref{table:msd_stats}. They are plotted with 95\% confidential interval computed by bootstrapping \cite{mooney1993bootstrapping}.}
	\label{fig:subset_errorbars_a}
\end{figure}

\begin{figure}[t]
	\centering
	\includegraphics[width=1.0\linewidth]{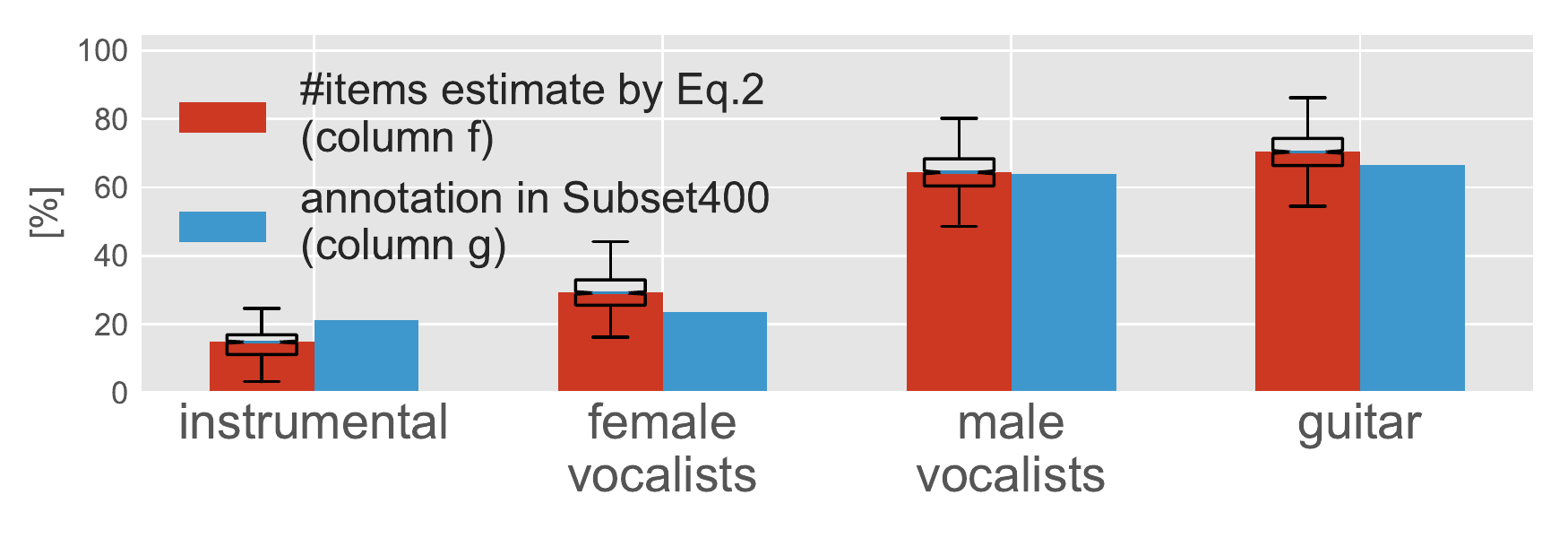}
	\caption{The estimates of the number of items (red) and the number of items in Subset400 (blue), both in percentage (they correspond to columns (f) and (g) in Table \ref{table:msd_stats}). The estimates are plotted with 95\% confidential interval computed by bootstrapping \cite{mooney1993bootstrapping}.}
	\label{fig:subset_errorbars_b}
\end{figure}

\subsubsection{Measuring label noise and tagability}\label{sssec:groundtruth_val}
Table \ref{table:msd_stats} column (a)-(d) summarises the statistics of Subset100. Confidence intervals for precision and recall are computed by bootstrapping \cite{mooney1993bootstrapping} and plotted in Figure \ref{fig:subset_errorbars_a}. The average error rate of negative labels is 42.5\%, which is very high, while that of positive labels is 3.5\%. As a result, the precision of the groundtruth is high (96.5\% on average) while the recall is much lower (71.9\% on average). This suggests that the tagging problem should be considered weakly supervised learning to some extent. We expect this problem exists in other weakly-labelled datasets as well, since annotators typically do not utilise all possible labels.

Such a high error rate for negative labels suggests that the tag occurrence counts in the groundtruth are under-represented. This can be related to the \textit{tagability} of labels, a notion which may be defined as \textit{the likelihood that a track will be tagged positive for a label when it really is positive}. If the likelihood is replaced with the portion of items, tagability is measured by recall as presented in Table \ref{table:msd_stats} as well as in Figure \ref{fig:subset_errorbars_a}. For example, bass guitar is one of the most widely used instruments in modern popular music, but it is only the 238\ts{th} most popular tag in the MSD since tagging music with `bass guitar' does not provide much information from the perspective of the average listener. Given the scores, we may assume that `female vocalists' (88.7\% of recall) and `instrumental' (80.0\%) are more tagable than `male vocalists' (60.5\%) and `guitar' (58.3\%), which indicates that the latter are presumably considered less unusual. 

\begin{figure}[t]
	\centering
	\includegraphics[width=1.0\linewidth]{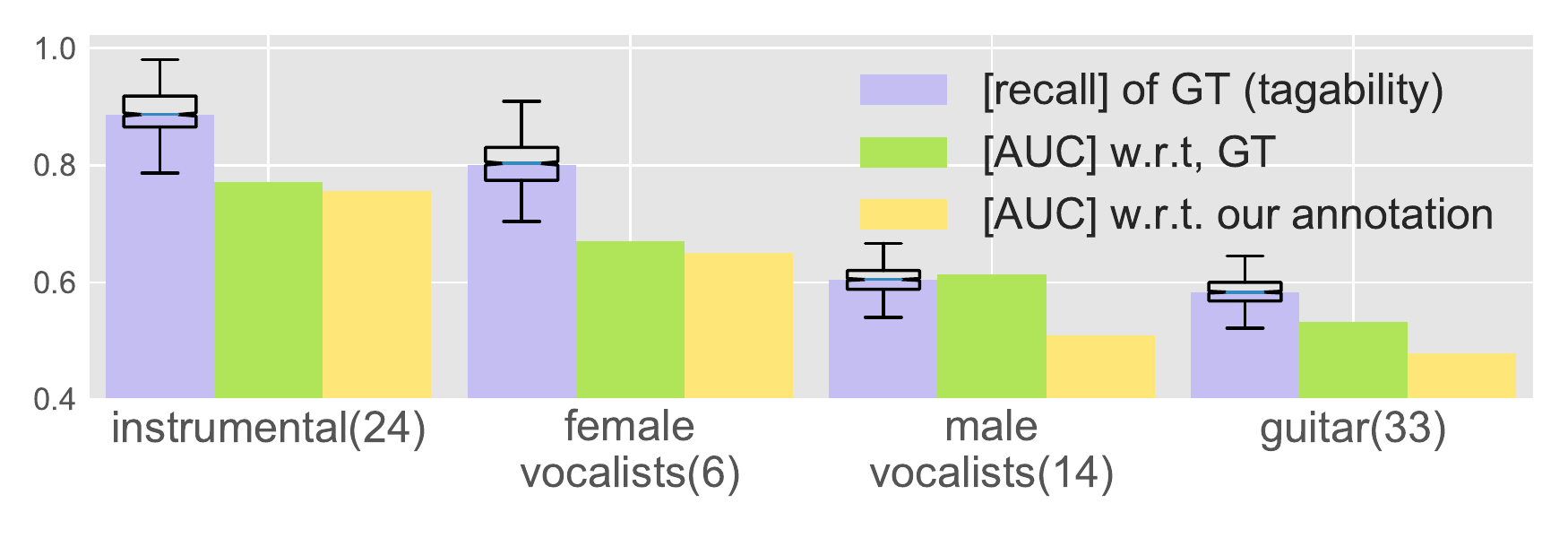}
	\caption{The recall rates (tagability, pink), AUC scores with respect to the groundtruth (green), and AUC scores with respect to our annotation (yellow), all reported on Subset400. The numbers on the x-axis labels are the corresponding popularity rankings of tags out of 50. The recall rates and their 95\% confidential intervals are identical to Figure \ref{fig:subset_errorbars_a} but plotted again for comparison with tag-wise AUC scores.}
	\label{fig:figssubset400withtaggability}
\end{figure}

\begin{figure*}
	\centering
	\includegraphics[width=1.0\linewidth]{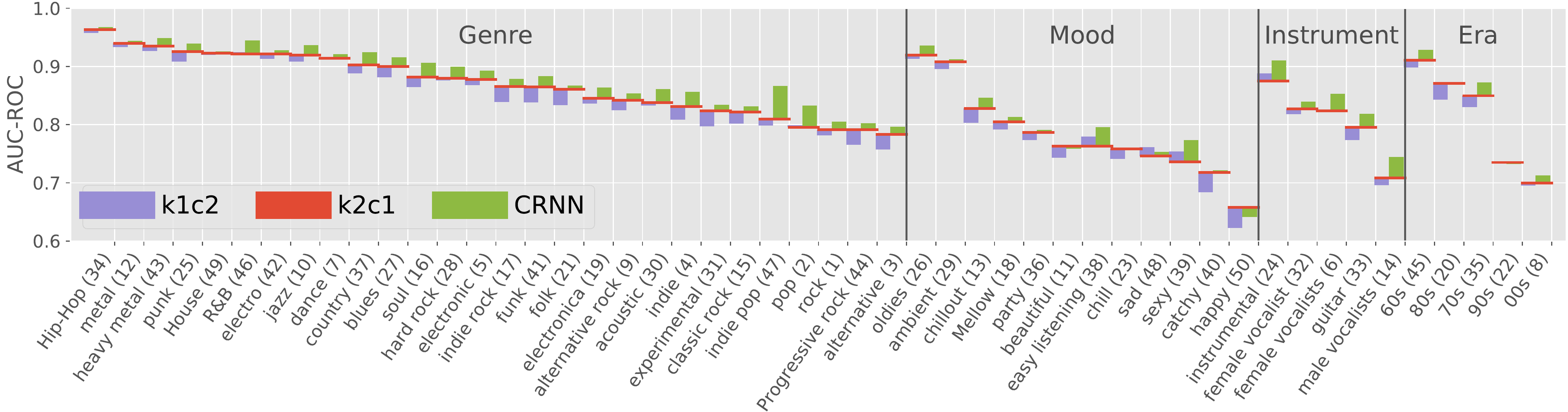}
	\caption{k1c2, k2c1, and CRNN are the names of network structures in \cite{choi2017convolutional}. \textit{i}) AUC of each tag is plotted using a bar chart and line. For each tag, the red line indicates the score of \texttt{k2c1} which is used as a baseline of bar charts for \texttt{k1c2} (blue) and \texttt{CRNN} (green). In other words, the blue and green bar heights represent the performance gaps, \texttt{k2c1}-\texttt{k1c2} and \texttt{CRNN}-\texttt{k2c1}, respectively. \textit{ii}) Tags are grouped by categories (genre/mood/instrument/era) and sorted by the score of \texttt{k2c1}. \textit{iii}) The number in parentheses after each tag indicates that tag's popularity ranking in the dataset.}
	\label{fig:figsaucpertags1m}
    \vspace{-1pt}
\end{figure*}

The correct number of \textit{positive}\hspace{0.03cm}/\hspace{0.03cm}\textit{negative} items can be estimated by applying Bayes' rule with the error rate. The estimated positive label count $\hat{N}^+$ is calculated using Eq.\ref{eq:est_count} as follows:
\begin{equation} \label{eq:est_count}
\hat{N}^+=N^+(1-p^{+}) + (T - N^+)p^-,
\end{equation} 
where $N^+$ is the tag occurrence, T is the number of total items ($T=242,842$ in our case), and $p^{+}$, $p^{-}$ refers to the error rates of positive and negative labels respectively. Column (f) of Table \ref{table:msd_stats} and Figure \ref{fig:subset_errorbars_b} present the estimated occurrence counts using Equation \ref{eq:est_count}. This estimate is validated using Subset400. Comparing the percentages in columns (f) and (g) confirms that the estimated counts are more correct than the tag occurrences of the groundtruth. For all four tags, the confidence intervals overlap with the percentage counted in Subset400 as illustrated in Figure \ref{fig:subset_errorbars_b}.
In short, the correct occurrence count is not correlated with the occurrence in the dataset, which shows the bias introduced by tagability. For instance, `male vocalists' is more likely to occur in music than `female vocalists', which means it has lower tagability, and therefore it ends up having fewer occurrences in the groundtruth.

\subsubsection{Effects of incorrect groundtruth on the training}\label{sssec:fx_training}

Despite such inaccuracies, it is possible to train networks for tagging with good performances using the MSD, achieving an AUC between 0.85 \cite{choi2016automatic} and 0.90 \cite{lee2017multi}. This may be because even with such noise, the network is weakly supervised by stochastically correct feedbacks, where the noise is alleviated by a large number of training examples \cite{torresani2014weakly}. In other words, given $\mathbf{x}$ is the input and $\mathbf{y}_{true}$, $\mathbf{y}_{noisy}$ are the correct and noisy labels respectively, the network can approximate the relationship $f: \mathbf{x} \rightarrow \mathbf{y}_{true}$ when training using $(\mathbf{x}, \mathbf{y}_{noisy})$. 

However, we suggest that the noise affects the training and it is reflected in the performances of different tags. In \cite{choi2017convolutional}, where different deep neural network structures for music tagging were compared, the authors observed a pattern on the per-tag performances that is common among different structures. This is illustrated in Figure \ref{fig:figsaucpertags1m} where the x-axis labels represent tag popularity ranking. The performances were not correlated with the ranking, the reported correlation is 0.077, therefore the question remained unanswered in \cite{choi2017convolutional}.
	
We conjecture that tagability, which is related to (negative) label noise can explain tag-wise performance differences. It is obvious that a low tagability implies more false negatives in the groundtruth. Therefore we end up feeding the network with more confusing training examples. For example, assuming there is a pattern related to male vocalists, the positive-labelled tracks provide mostly correct examples. However, many examples of negative-labelled tracks (64\% in Subset100) also exhibit the pattern. Consequently, the network is forced to distinguish hypothetical differences between the positive-labelled true patterns and the negative-labelled true patterns, which leads to learning a more fragmented mapping of the input. This is particularly true in music tagging datasets where negative label noise dominates the total noise. This is supported by data both in this paper and \cite{choi2017convolutional} as discussed below.

First, tagabilities (or recall) and AUC scores with respect to the groundtruth and our re-annotation are plotted in Figure \ref{fig:figssubset400withtaggability} using Subset400 items and the \texttt{compact-convnet} structure. Both AUC scores are positively correlated to tagability while they are not related to the tag popularity rankings. Although the confidence intervals of `instrument' vs. `female vocalists', and `male vocalists' vs. `guitar' overlap, there is an obvious correlation. The performances on the whole test set also largely agree with our conjecture. 

Second, in Figure \ref{fig:figsaucpertags1m}, AUC scores for instrument tags are ranked as `instrumental' $>$ `female vocalists' $>$ `guitar' $>$ `male vocalists' for all three ConvNet structures. This aligns with tagability in Figure~\ref{fig:figssubset400withtaggability} within the confidence intervals.

This observation motivates us to expand this approach and assess tags in other categories. Within the Era category in Figure \ref{fig:figsaucpertags1m}, performance is negatively correlated with the popularity ranking (Spearman correlation coefficient=-0.7). There is a large performance gap between old music groups (60s, 80s, 70s) and the others (90s, 00s). We argue that this may also be due to tagability. In the MSD, older songs (e.g. those released before the 90s) are less frequent compared to modern songs (90s or 00s). According to the year prediction subset of the MSD\footnote{\url{https://labrosa.ee.columbia.edu/millionsong/pages/tasks-demos}}, 84\% of tracks are released after 1990. This is also related to the fact that the tag `oldies' exists while its opposite does not. Hence, old eras seem more tagable, which might explain the performance differences in Era tags. We cannot extend this approach to mood and genre tags because the numbers of tags are much larger and there may be aspects contributing to tag-wise performance differences other than tagability.

\subsubsection{Validation of the evaluation}\label{sssec:val_eval}
Another problem with using a noisy dataset is evaluation. In the previous section, we assumed that the system can learn a \textit{denoised} relationship between music pieces and tags, $f: \mathbf{x} \rightarrow \mathbf{y}_{true}$. However, the evaluation of a network with respect to $\mathbf{y}_{noisy}$ includes errors due to noisy groundtruth. This raises the question of the reliability of the results. We use our strongly-labelled annotation of Subset400 to assess this reliability.

\begin{table*}[t]
	\centering
	\caption{Top-20 Similar tag tuples by two analysis approaches. The first row is by analysing co-occurrence of tags in groundtruth (see \ref{ssec:msd_tags} for details). The second row is by the similarity of trained label vector (see \ref{ssec:label_vector} for details). Common tuples are annotated with matching symbols.}
	\label{table:label_pairs}
	\begin{tabular}{lll}
		\cline{1-2}
		\multicolumn{1}{|c|}{\begin{tabular}[c]{@{}c@{}}Similar tags by \\groundtruth labels\end{tabular}}  & \multicolumn{1}{l|}{\begin{tabular}[c]{@{}l@{}}(alternative rock, rock)\textsuperscript{$\mathsection$} (indie rock, indie)\textsuperscript{\#} (House, dance)\textsuperscript{$\ddagger\ddagger$} (indie pop, indie) (classic rock, rock)\\(electronica, electronic)\textsuperscript{*} (alternative, rock) (hard rock, rock) (electro, electronic)\textsuperscript{**} (House, electronic)\\(alternative rock, alternative)\textsuperscript{$\mathparagraph$} (catchy, pop) (indie rock, rock) (60s, oldies)\textsuperscript{$\dagger\dagger$} (heavy metal, metal)\textsuperscript{$\mathsection\mathsection$}\\(rnb, soul) (ambient, electronic) (90s, rock) (heavy metal, hard rock)\textsuperscript{$\ddagger$} (alternative, indie)\textsuperscript{$\|$}\end{tabular}}                                                                                       &  \\ \cline{1-2}
		\multicolumn{1}{|c|}{\begin{tabular}[c]{@{}c@{}}Similar tags by \\label vectors\end{tabular}} & \multicolumn{1}{l|}{\begin{tabular}[c]{@{}l@{}}(electronica, electronic)\textsuperscript{*} (indie rock, indie)\textsuperscript{\#} (female vocalist, female vocalists) (heavy metal, hard rock)\textsuperscript{$\ddagger$} (indie, indie pop)\\ (sad, beautiful) (alternative rock, rock)\textsuperscript{$\mathsection$} (alternative rock, alternative)\textsuperscript{$\mathparagraph$} (happy, catchy) (indie rock, alternative)\\ (alternative, indie)\textsuperscript{$\|$} (rnb, sexy) (electro, electronic)\textsuperscript{**} (sad, Mellow) (Mellow, beautiful)\\ (60s, oldies)\textsuperscript{$\dagger\dagger$} (House, dance)\textsuperscript{$\ddagger\ddagger$} (heavy metal, metal)\textsuperscript{$\mathsection\mathsection$} (chillout, chill) (electro, electronica)\end{tabular}} &  \\ \cline{1-2}
		\textsl{}
	\end{tabular}
\end{table*}

Let us re-examine Figure \ref{fig:figssubset400withtaggability}. All AUC scores with respect to our annotation are lower than the scores with respect to the groundtruth. Performance for the \textit{guitar} tag is remarkably below 0.5, the baseline score of a random classifier. However, the overall trend of tag-wise performance does not change. Because the results are based only on four classes and a subset of songs, a question arises: How does this result generalise to other tags?

\begin{figure}[t]
	\centering
	\includegraphics[width=1.0\columnwidth]{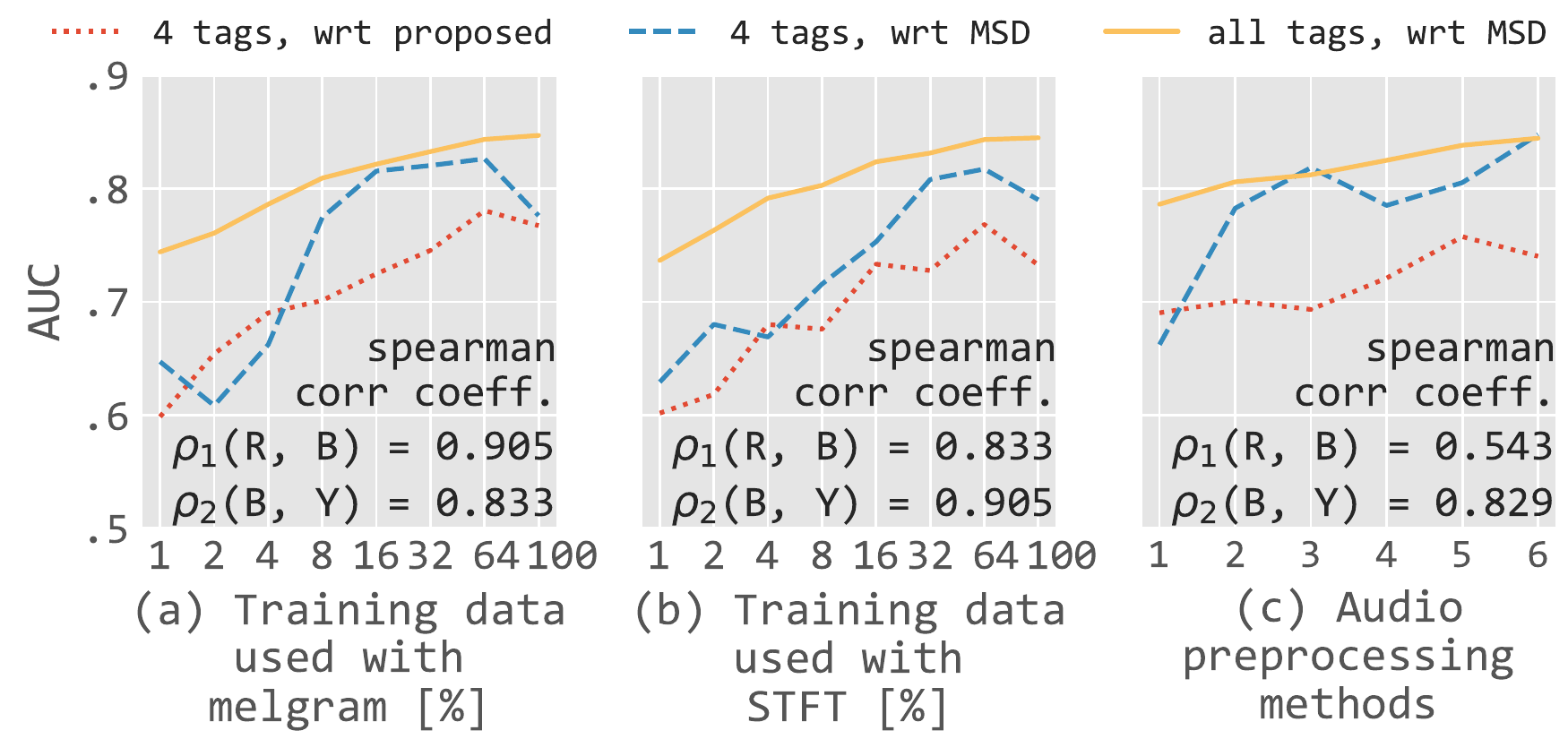} 
	\caption{AUC scores of all tags (yellow, solid) and four instrumentation tags. Instrumentation tags are evaluated using \textit{i)} dataset groundtruth (blue, dashed) and \textit{ii)} our strong-labelling re-annotation (red, dotted). Pearson correlation coefficients between \{red vs. blue\} and \{blue vs. yellow\} is annotated on each chart as $\rho$. In (c), x-axis is experiment index and various audio preprocessing methods were applied for each experiment.}
	\label{fig:ds_validation}
\end{figure}

To answer the question, three AUC scores are plotted in Figure \ref{fig:ds_validation}: \textit{i)} the scores of the four instrument tags with respect to our annotation (dotted red), 
\textit{ii)} the scores of the four instrument tags with respect to the given groundtruth (dashed blue), and \textit{iii)} the scores of all tags with respect to the given groundtruth (solid yellow).

The reliability of evaluation is typically assessed with a given groundtruth and can be measured by $\rho_1$, the Pearson correlation coefficient between AUCs using our annotation and the MSD. The correlation between the four tags and all other tags (shown in blue and yellow), denoted $\rho_2$, is a measure of how we can generalise our re-annotation result to those concerning all tags.

We selected three sets of tagging results to examine and plotted these in Figure \ref{fig:ds_validation} (a-c). The first two sets shown in subfigure (a) and (b) are results after training the \texttt{compact-convnet} with varying training data size and different audio representations: (a) melspectrogram and (b) short-time Fourier transform. The third set of curves in Figure \ref{fig:ds_validation} (c) compare six results with varying input time-frequency representations, training data size and input preprocessing techniques including normalisation and spectral whitening. The third set is selected to observe the correlations when the performance differences among systems are more subtle than those in (a) and (b).\footnote{We omit the details of the preprocessing methods, which are summarised in \cite{choi2017comparison}, because the focus is on the correlation of the final scores.}
First, the $\rho_1$ values in (a)--(c) suggest that noisy groundtruth provides reasonably stable evaluation for the four instrument tags. On the first two sets in (a) and (b), the scores of four tags using the MSD groundtruth (in blue) are highly correlated ($\rho_1=0.905$ and $0.833$) to the scores using our annotation (red). This suggests the evaluation using noisy labels is still reliable. However, in (c), where the scores of all tags with given groundtruth (yellow) are in a smaller range, the correlation between all tags and the four tags ($\rho_1$) decreases to 0.543. The results imply that the distortion on the evaluation using the noisy groundtruth may disguise the performance difference between systems when the difference is small. Second, large $\rho_2$ indicates that our validation is not limited to the four instrument tags but can be generalised to all tags. The correlation coefficients $\rho_2$ is stable and reasonably high in (a)--(c). It is 0.856 on average.

\subsection{Analysis of predicted label vectors} \label{ssec:label_vector}
In the previous sections, groundtruth labels were analysed from various perspectives. It is worth considering how this information is distilled into the network after training, and whether we can leverage the trained network beyond our particular task. To answer these questions we use the trained network weights to assess how the network `understands' music content by its label. This analysis also provides a way to discover unidentified relationships between labels and the music. The goal of label vector analysis is to better understand network training as well as assess its capacity to represent domain knowledge, i.e., relationships between music tags that are not explicitly shown in the data. 
In the \texttt{compact-convnet} described in Section \ref{sec:dataset}, the output layer has a dense connection to the last convolutional layer. The weights are represented as a matrix $\textbf{W} \in \mathbb{R}^{N\times 50}$, where $N$ is the number of feature maps ($N$=32 for our case) and 50 is the number of the predicted labels. After training, the columns of $\textbf{W}$ can be interpreted as $N$-dimensional latent vectors since they represent how the networks combine information in the last convolutional layer to make the final prediction. We call these \textit{label vector}s.

We compute the pairwise label vector similarity (LVS) using the dot product, i.e., $S(i, j) = w(i) \cdot w(j)$ where $i, j \leq 50 $ or equivalently:
\begin{equation} \label{eq:lvsm}
\textbf{S} = \textbf{W}^\top \cdot \textbf{W},
\end{equation}
which yields a $50$$\times$$50$ symmetric matrix.

\begin{figure}[t]
	\centering
	\includegraphics[width=1.0\columnwidth]{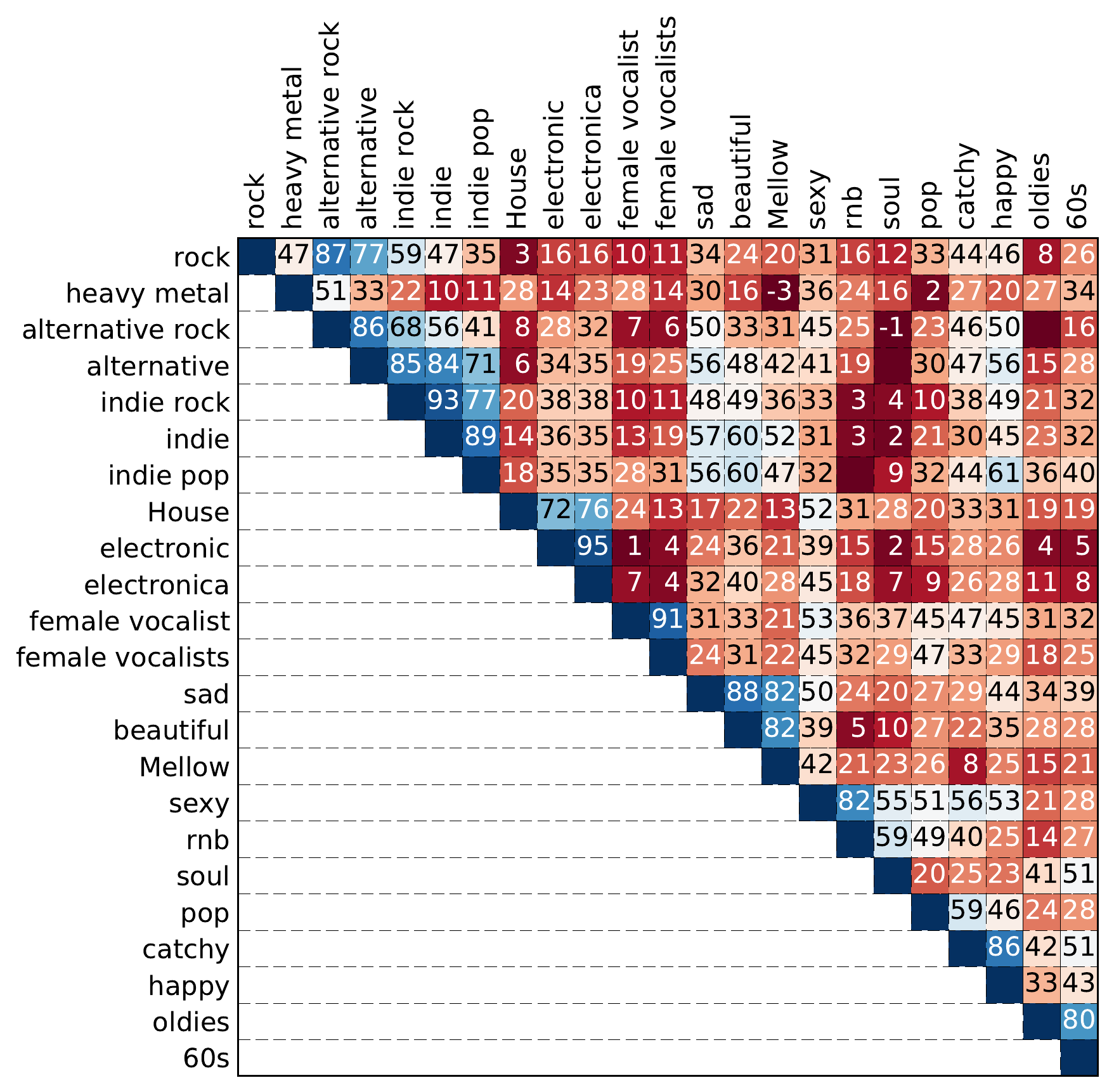}
	\caption{Label vector similarity matrix by Eq. \ref{eq:lvsm} (of manually selected 23 tags, same in Figure \ref{fig:tag_cooccur}, where symmetric components are omitted and numbers are $\times100$ after dot product for visual clarity.}
	\label{fig:label_sim}
\end{figure}

LVS is illustrated in Figure \ref{fig:label_sim}. The pattern is similar to the values in NCO (normalised co-occurrence) shown in Figure \ref{fig:tag_cooccur} (see Sec.\ref{ssec:labelling_strategies}). On average, the similarities in $S(i, j)$ are higher than those in $C(i, j)$. In $\textbf{S}$, only four pairs show negative values, `classic rock' -- `female vocalists', and `mellow' -- \{`heavy metal', `hard rock', and `dance'\}. In other words, label vectors are distributed in a limited space corresponding to a 32 dimensional vector space, where the angle $\theta$ between $w(i)$ and $w(j)$ is smaller than $\pi/2$ for most of the label vector pairs.
This result can be interpreted in two ways: how well the ConvNet reproduce the co-occurrence that was provided by the training set; and if there is additional insight about music tags in the trained network.

First, the Pearson correlation coefficient of the rankings by LVS and NCO is 0.580.\footnote{Because of the asymmetry of $C(i, j)$, rankings of $\max (C(i, j), C(y, y))$ are used.}. The top 20 most similar label pairs are sorted and described in Table \ref{table:label_pairs}. The second row of the table shows similar pairs according to the label vectors estimated by the network. Eleven out of 20 pairs overlap with the top 20 NCO tuples shown in the top row of the table. Most of these relations can be explained by considering the genre hierarchy. Besides, pairs such as (`female vocalists', `female vocalists') and (`chillout', `chill') correspond to semantically similar words. Overall, tag pairs showing high similarity (LVS) reasonably represent musical knowledge and correspond to high NCO values computed from the ground truth. This confirms the effectiveness of the network to predict subjective and high-level semantic descriptors from audio only.

Second, there are several pairs that are \textit{i)} high in LVS, \textit{ii)} low in NCO, and \textit{iii)} presumably music listeners would reasonably agree with their high similarity. These pairs show the extracted representations of the trained network can be used to measure tag-level musical similarities even if they are not explicitly shown in the groundtruth. For example, pairs such as (`sad', `beautiful'), (`happy', `catchy') and (`rnb', `sexy') are in the top 20 of LVS (6\ts{th}, 9\ts{th}, and 12\ts{th} similarities with 0.88, 0.86, and 0.82 of similarity values respectively). On the contrary, according to the ground truth, they are only 129\ts{th}, 232\ts{nd}, 111\ts{th} co-occurring with 0.13, 0.08, and 0.14 of co-occurrence likelihood respectively.

In summary, the analysis based on LVS indirectly validates that the network learned meaningful representations that correspond to the groundtruth. Moreover, we found several pairs that are considered similar by the network which may help to extend our understanding of the relation between music and tags.


\section{Conclusions}\label{sec:conclusion}
In this article, we investigated several aspects how noisy labels in folksonomies affect the training and performance of deep convolutional neural networks for music tagging. We analysed the MSD, the largest dataset available for training a music tagger from a novel perspective. We reported on a study aiming to validate the MSD as groundtruth for this task. We found that the dataset is reliable overall, despite several noise sources affecting training and evaluation. Finally, we defined and used \textit{label vectors} to analyse the capacity of the network to explain similarity relations between semantic tags.

Overall, the behaviours of the trained network were shown to be related to the property of the given labels. The analysis showed that tagability, which we measured by recall on the groundtruth, is correlated to the tag-wise performance. This opened a way to explain tag-wise performance differences within other categories of tags such as era. In the analysis of the trained network, we found that the network learns more intricate relationships between tags rather than simply reproducing the co-occurrence patterns in the groundtruth. The trained network is able to infer musically meaningful relationships between tags that are not present in the training data. 

Although we focused on music tagging, our results provide general knowledge applicable in several other domains or tasks including other music classification tasks. The analysis method presented here and the result on the tagging dataset can easily generalise to similar tasks in other domains involving folksonomies with noisy labels or tasks involving weakly labelled datasets, e.g. image tagging, object recognition in video, or environmental sound recognition, where not all sources are necessarily labelled. Future work will explore advanced methods to learn and evaluate using noisy datasets
under a structured machine learning framework. Tagability can be understood from the perspective in music cognition research and should be investigated further. 

\section*{Acknowledgements}
This work is supported by EPSRC project (EP/L019981/1) `Fusing Semantic and Audio Technologies for Intelligent Music Production and Consumption' and the European Commission H2020 research and innovation grant AudioCommons (688382). Sandler acknowledges the support of the Royal Society as a recipient of a Wolfson Research Merit Award. Choi acknowledges the support of QMUL Postgraduate Research Fund for research visiting to NYU. Cho acknowledges eBay, TenCent, Facebook, Google and NVIDIA. 

\ifCLASSOPTIONcaptionsoff
  \newpage
\fi

\bibliographystyle{IEEEtran}
\bibliography{convnets}

%

\begin{IEEEbiography}[{\includegraphics[width=1in,height=1.25in,clip,keepaspectratio]{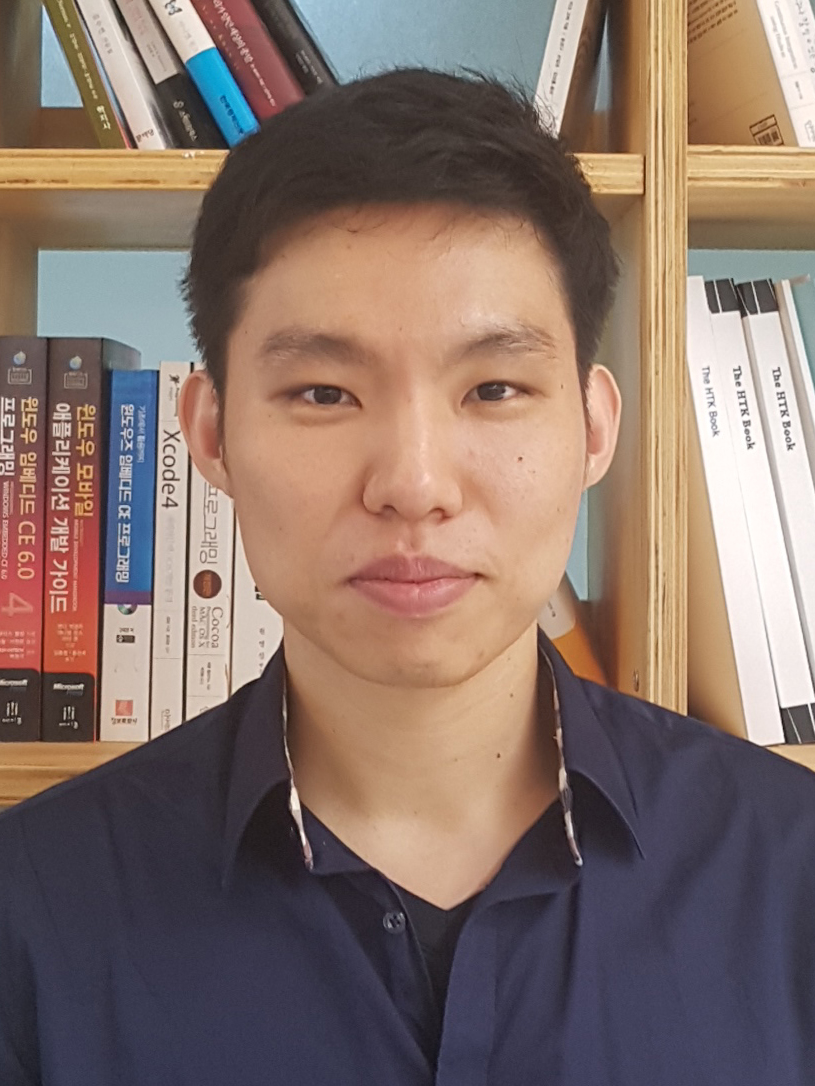}}]{Keunwoo Choi}
received his B.Sc and M.Phil degrees in electrical engineering and computer science from Seoul National University, Seoul, South Korea, in 2009 and 2011, respectively. In 2011, He joined Electrical and Telecommunication Research Institute, Daejeon, South Korea, as a researcher. He is a PhD candidate at the Centre for Digital Music (C4DM), School of Electronic Engineering and Computer Science in Queen Mary University of London, London, UK, in 2014. 
\end{IEEEbiography}

\begin{IEEEbiography}[{\includegraphics[width=1in,height=1.25in,clip,keepaspectratio]{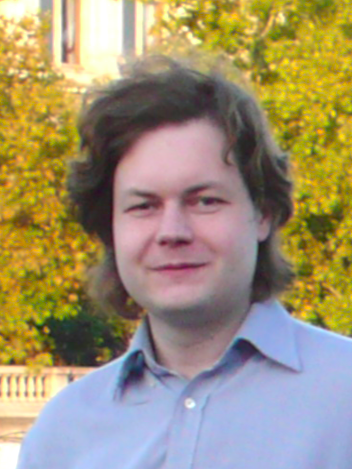}}]{Gy\"orgy Fazekas}
is a lecturer at Queen Mary University of London (QMUL), working at the Centre for Digital Music (C4DM), School of Electronic Engineering and Computer Science. He received his BSc degree at Kandó Kálmán College of Electrical Engineering, Faculty of Electrical Engineering, Óbuda University, Budapest, Hungary. He received his MSc and PhD degrees at QMUL, United Kingdom in 2012. His thesis titled “Semantic Audio Analysis-Utilities and Applications” explores novel applications of semantic audio analysis, semantic web technologies and ontology based information management. His research interests include the development of semantic audio technologies and their application to creative music production. He is leading QMUL’s research on the EU funded Audio Commons project facilitating the use of open sound content in professional audio production. He collaborates on several other projects and he is a member of the IEEE, AES and ACM. 
\end{IEEEbiography}

\begin{IEEEbiography}[{\includegraphics[width=1in,height=1.25in,clip,keepaspectratio]{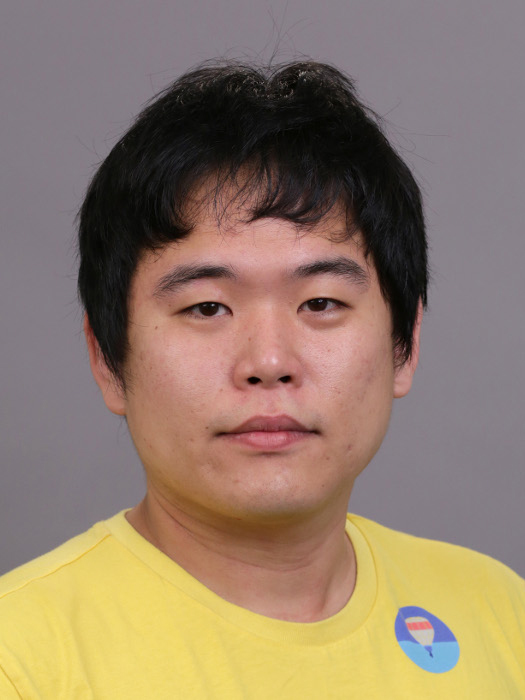}}]
{Kyunghyun Cho} is an assistant professor of computer science and data science at New York University. He was a postdoctoral fellow at University of Montreal until summer 2015 under the supervision of Prof. Yoshua Bengio, and received PhD and MSc degrees from Aalto University early 2014 under the supervision of Prof. Juha Karhunen, Dr. Tapani Raiko and Dr. Alexander Ilin. 
\end{IEEEbiography}

\begin{IEEEbiography}[{\includegraphics[width=1in,height=1.25in,clip,keepaspectratio]{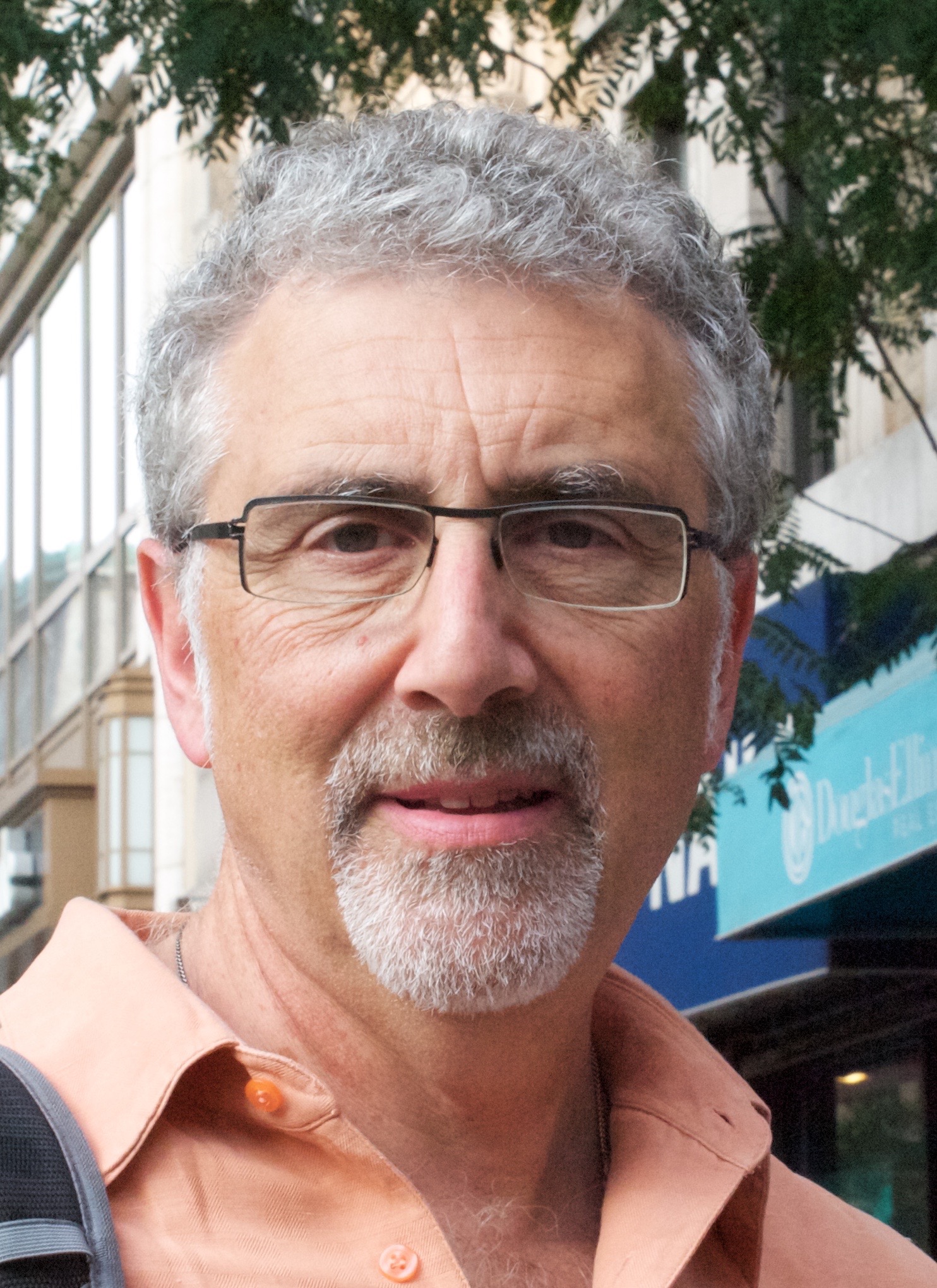}}]{Mark Sandler}
 was born in 1955. He received the BSc and PhD degrees from the University of Essex, UK, in 1978 and 1984 respectively. He is Professor of Signal Processing and Founding Director of the Centre for Digital Music in the School of Electronic Engineering and Computer Science at Queen Mary University of London, UK. He has published over 400 papers in journals and conferences. He is Fellow of the Royal Academy of Engineering, IEEE, AES and IET. \end{IEEEbiography}

%
%
%




\end{document}